\documentclass[doublecol]{epl2}

\title{Computer simulation of Wheeler's delayed choice experiment with photons\footnote{Accepted for publication in Europhysics Letters}}
\shorttitle{Computer simulation of Wheeler's delayed choice experiment} 

\author{S. Zhao\inst{1} \and S. Yuan\inst{1} \and H. De Raedt\inst{1} \and K. Michielsen\inst{2}}
\shortauthor{Shuang Zhao \etal}

\institute{
  \inst{1}Department of Applied Physics, Zernike Institute for Advanced Materials,
University of Groningen, Nijenborgh 4, NL-9747 AG Groningen, The Netherlands\\
  \inst{2} EMBD, Vlasakker 21, B-2160 Wommelgem, Belgium
}

\def\revision#1{#1}

\abstract{
We present a computer simulation model of Wheeler's delayed choice experiment
that is a one-to-one copy of an experiment reported recently (V. Jacques {\sl et al.}, Science 315, 966 (2007)).
The model is solely based on experimental facts,
satisfies Einstein's criterion of local causality and does not rely on any concept of quantum theory.
Nevertheless, the simulation model reproduces the averages as obtained from
the quantum theoretical description of Wheeler's delayed choice experiment.
Our results prove that it is possible to give a particle-only description of Wheeler's delayed choice experiment
which reproduces the averages calculated from quantum theory and which does not defy common sense.
}

\begin{document}

\maketitle

\section{Introduction}

According to the wave-particle duality, a concept of quantum theory, photons
exhibit both wave and particle behavior depending upon the circumstances of
the experiment~\cite{HOME97}. In 1978, Wheeler proposed a gedanken
experiment~\cite{WHEE83}, a variation on Young's double slit experiment, in which the
decision to observe wave or particle behavior is made after the photon has
passed the slits. The pictorial description of this experiment defies common sense:
The behavior of the photon in the past is said to be influenced changing the representation of the photon from a particle to a wave.

\begin{figure}[t]
\begin{center}
\includegraphics[width=8.5cm]{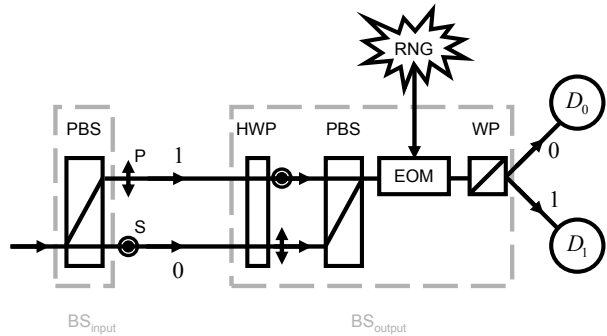}
\caption{Schematic diagram of the experimental setup for Wheeler's delayed-choice gedanken
experiment~\cite{JACQ07}. PBS: Polarizing beam splitter; HWP:
Half-wave plate; EOM: electro-optic modulator; RNG: Random number generator;
WP: Wollaston prism; P,S: Polarization state of the photon; $D_{0}$, $D_{1}$: Detectors.}
\label{wheeler}
\end{center}
\end{figure}

Recently, Jacques {\sl et al.} reported an almost ideal
experimental realization of Wheeler's delayed choice experiment~\cite{JACQ07}.
The experimental set-up (see Fig.~\ref{wheeler}) consists of a
single-photon source, a Mach-Zehnder interferometer, with at the output side
a beam splitter (BS$_{output}$) of which the presence can be controlled by a
voltage applied to an electro-optic modulator (EOM) and detectors~\cite{JACQ07}.
\revision{Although the detection events are the only experimental facts,
the pictorial description of Jacques {\sl et al.}~\cite{JACQ07} is as follows:
The decision to apply a voltage to the EOM is made after the photon
has passed BS$_{input}$ but before the photon
enters BS$_{output}$.}
If no voltage is applied to the EOM (open configuration), then the arrival of a photon at
either detector clearly gives which-way information about the photon within
the interferometer (particle behavior), with 50\% arriving from either path.
When the voltage is applied (closed configuration), the paths interfere and
it is impossible to know which path the photon took (wave behavior).
Accordingly, the detectors register an interference pattern.

The outcome of delayed-choice experiments~\cite{HELL87,BALD89,LAWS96,KAWA98,KIM00,JACQ07}, that is the average results of many detection
events, is in agreement with quantum theory.
However, the pictorial description~\cite{JACQ07} defies common sense: The decision to apply a voltage
to the EOM after the photon left BS$_{input}$ but before it passes BS$_{output}$,
influences the behavior of the photon in the past and changes the
representation of the photon from a particle to a wave~\cite{JACQ07}. \revision{On
the other hand, quantum theory does not describe single events~\cite{HOME97}.}
Therefore, it should not be a surprise that the application of concepts
of quantum theory to the domain of individual events may lead to conclusions
that are at odds with common sense.

In this Letter, we describe a model that, when implemented
as a computer program, performs an event-by-event simulation
of Wheeler's delayed-choice experiment.
Every essential component of the laboratory experiment
(PBS, EOM, HWP, Wollaston prism, detector) has a counterpart in the
algorithm.
The data is analyzed by counting detection events, just like in the experiment~\cite{JACQ07}.
The simulation model is solely based on experimental facts, satisfies Einstein's criterion
of local causality and does not rely on any concept of quantum theory \revision{or of probability theory}.
Nevertheless, our simulation model reproduces the averages obtained from the quantum theoretical
description of Wheeler's delayed choice experiment but as our approach does
not rely on concepts of quantum theory and gives a description on the
level of individual events, it provides a description of the experimental facts that
does not defy common sense.
\revision{In a pictorial description of our simulation model, we may speak about ``photons'' generating
the detection events. However, these so-called photons, as we will call them in the sequel,
are  elements of a model or theory for the real laboratory experiment only.
The experimental facts are the settings of the various apparatuses and
the detection events. What happens in between activating the source and the registration of the detection
events is not measured and is therefore not known.}
Although the \revision{photons} ``know'' exactly which route they followed
in the closed configuration of the interferometer (we can
always track them during the simulation), they build up an interference
pattern at the detector. The appearance of an interference pattern is
commonly considered to be characteristic for a wave. In this Letter, we demonstrate
that, as in experiment, it can also be build up by many \revision{photons}.
These \revision{photons} have which-way information, never directly communicate with each other
and arrive one by one at a detector.

To head off possible misunderstandings, the work presented here is not concerned with the interpretation or an extension of quantum theory.
We adopt the point of view that quantum theory has nothing to say about individual events~\cite{HOME97}.
The fact that there exist simulation algorithms that reproduce the results of quantum theory
has no direct implications to the foundations of quantum theory:
These algorithms describe the process of generating events at a level of detail
that is outside the scope of what current quantum theory can describe.
The event-based simulation approach that we describe in this Letter
is unconventional in that it does not require knowledge of the probability distribution
obtained by solving the quantum problem.
\revision{The averages given by quantum theory are obtained through a simulation of locally causal,
classical dynamical systems.}
The key point of these dynamical systems is that they are built from units that are adaptive.

\revision{It is common practice to use the framework of Kolmogorov 's probability theory to construct
probabilistic models of phenomena that cannot (yet) be described by a deductive theory. Although Kolmogorov 's
probability theory provides a rigorous framework to formulate such models, there are ample examples that
illustrate how easy it is to make plausible assumptions that create all kinds of paradoxes,
also for every-day problems that have no bearing on quantum theory at all~\cite{GRIM95,TRIB69,JAYN03,HONE02,BALL03}.
Subtle mistakes such as dropping (some of the) conditions, mixing up the meaning of physical and statistical independence,
and changing one probability space for another during the cause of an argument can give rise to
all kinds of paradoxes~\cite{JAYN89,HESS01,HESS06}.
To avoid these potential pitfalls, in our simulation approach we strictly stay
in the domain of integer arithmetic, that is we do not invoke any concept of probability theory.}

This Letter builds on earlier work~\cite{RAED05d,RAED05b,RAED05c,MICH05,RAED06c,RAED07a,RAED07b,RAED07c,ZHAO08} that
demonstrated that it may be possible to simulate quantum phenomena on the level of individual events
without invoking a single concept of quantum theory.
Specifically, we have demonstrated that locally-connected networks of processing units
with a primitive learning capability can simulate event-by-event,
the single-photon beam splitter and Mach-Zehnder interferometer experiments of Grangier {\sl et al.}~\cite{GRAN86}
and Einstein-Podolsky-Rosen experiments with photons~\cite{ASPE82a,ASPE82b,WEIH98}.
Furthermore, we have shown that this approach can be generalized
to simulate universal quantum computation by an event-by-event process~\cite{MICH05,RAED05c}.
\revision{The algorithms used in our earlier work~\cite{RAED05d,RAED05b,RAED05c,MICH05,RAED06c,RAED07a,RAED07b,RAED07c,ZHAO08}
cannot be used to simulate Wheeler's delayed choice experiment~\cite{JACQ07}.
The latter uses components that respond to the polarization and/or the difference in path length
and the algorithms used in our earlier work cannot handle both features simultaneously.
In contrast, the more general algorithms described in this Letter can also simulate all the experiments
covered in our earlier work.}
Our event-by-event simulation approach rigorously satisfies
Einstein's criterion of local causality and builds up the final outcome that agrees
with quantum theory event-by-event, as observed in real experiments.

\section{Simulation model}

The simulation algorithm can be viewed as a message-processing and message-passing
process: It routes messengers through a network of units that process messages.
In a pictorial description, the photon is the messenger,
carrying a message representing its phase and polarization.
The processing units play the role of the components of the laboratory experiment
and the network represents the complete experimental setup.
Some processing units consist of an input stage (a standard linear adaptive filter),
a transformation stage and an output stage.
The input (output) stage may have several \revision{channels}
at (through) which messengers arrive (leave).
Other processing units are simpler in the sense that the input stage
is not necessary for the proper functioning of the device.
A message is represented by a vector.
As a messenger arrives at an input \revision{channel} of a processing unit,
the input stage updates its internal state, represented by a vector, and
sends the message together with its internal state
to the transformation stage that implements the operation of the particular device.
Then, a new message is sent to the output stage,
using a pseudo-random number to select the output \revision{channel}
through which the messenger will leave the unit.
\revision{We use pseudo-random numbers to mimic the apparent unpredictability of the
experimental data only.  The use of pseudo-random numbers is merely convenient, not essential.}
At any given time, there is only one messenger being routed through the whole network.
There is no direct communication between the messengers.

In the experimental realization of Wheeler's delayed choice experiment by
Jacques {\sl et al.}~\cite{JACQ07} linearly polarized single photons
are sent through a polarizing beam splitter (PBS) that
together with a second, movable PBS forms an interferometer (see Fig.~\ref{wheeler}).
The network of processing units is a one-to-one
image of the experimental setup~\cite{JACQ07} and is therefore not shown.
We now describe each of the components of the network in detail.

\subsection{Messenger}
Photons are regarded as messengers.
Each messenger carries a message represented by a
six-dimensional unit vector ${\bf y}_{k,n}=( \cos \psi
_{k,n}^{H},\sin \psi _{k,n}^{H},\cos \psi _{k,n}^{V},\sin \psi
_{k,n}^{V},\cos \xi _{k,n},\sin \xi _{k,n}) $.
The superscript H (V) refers to the horizontal (vertical)
component of the polarization
and $\psi _{k,n}^{H}$, $\psi _{k,n}^{V}$, and $\xi _{k,n}$
represent the phases and polarization of the photon, respectively.
It is evident that the representation used here maps one-to-one
to the plane-wave description of a classical electromagnetic field~\cite{BORN64},
except that we assign these properties to each individual message, not to a wave.
The subscript $n\geq 0$ numbers
the consecutive messages and $k=0,1$ labels the channel of the PBS
at which the message arrives (see below).

\begin{figure}[t]
\begin{center}
\includegraphics[width=8cm]{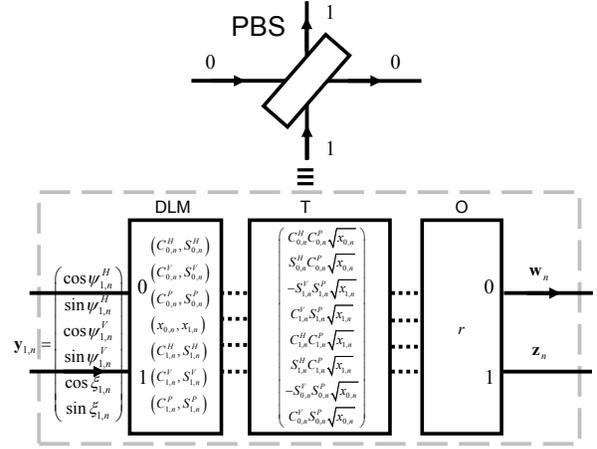}
\caption{Diagram of a DLM-based processing unit that performs an event-based simulation of a
polarizing beam splitter (PBS). The solid lines represent the
input and output channels of the PBS. The presence of a message is indicated
by an arrow on the corresponding channel line. The dashed lines indicate the
data flow within the PBS.}
\label{PBS}
\end{center}
\end{figure}

\subsection{Polarizing beam splitter}
The processor that performs the event-by-event simulation of a PBS is depicted in Fig.~\ref{PBS}.
It consists of an input stage, a simple deterministic learning machine (DLM)~\cite{RAED05d,RAED05b,RAED05c,MICH05},
a transformation stage (T), an output stage (O) and
has two input and two output channels labeled with $k=0,1$.
We now define the operation of each stage explicitly.

\begin{itemize}
\item{Input stage: The DLM receives a message on either input channel 0 or 1,
never on both channels simultaneously. The arrival of a message on channel 0
(1) is named a 0 (1) event. The input events are represented by the vectors
${\bf v}_{n}=(1,0)$ or ${\bf v}_{n}=(0,1)$ if the $n$th
event occurred on channel 0 or 1, respectively.
The DLM has six internal registers ${\bf Y}_{k,n}^{H}=(
C_{k,n}^{H},S_{k,n}^{H}) ,$ ${\bf Y}_{k,n}^{V}=(
C_{k,n}^{V},S_{k,n}^{V}) ,$ ${\bf Y}_{k,n}^{P}=(
C_{k,n}^{P},S_{k,n}^{P}) $ and one internal vector ${\bf x}%
_{n}=( x_{0,n},x_{1,n}) $, where $x_{0,n}+x_{1,n}=1$
and $x_{k,n}\geq 0$ for $k=0,1$ and all $n$. These seven two-dimensional vectors are labeled by the
message number $n$ because their contents are updated every time the DLM
receives a message.
Note that the DLM stores information about the last message only.
The information carried by earlier messages is overwritten by
updating the internal registers.

Upon receiving the $(n+1)$th input event, the DLM performs the following steps:
It stores the first two elements of message ${\bf y}_{k,n+1}$ in
its internal register ${\bf Y}_{k,n+1}^{H}=(
C_{k,n+1}^{H},S_{k,n+1}^{H}) $, the middle two elements
of ${\bf y}_{k,n+1}$ in
${\bf Y}_{k,n+1}^{V}=( C_{k,n+1}^{V},S_{k,n+1}^{V})$,
and the last two elements of ${\bf y}_{k,n+1}$
in
${\bf Y}_{k,n+1}^{P}=( C_{k,n+1}^{P},S_{k,n+1}^{P})$.
Then, it updates its internal vector according to the rule~\cite{RAED05d}
\begin{equation}
x_{i,n+1}=\alpha x_{i,n}+( 1-\alpha ) \delta _{i,k},
\end{equation}%
where $0<\alpha <1$ is a parameter that controls the learning process~\cite{RAED05d}.
Note that by construction $x_{0,n+1}+x_{1,n+1}=1$, $x_{0,n+1}\geq 0$ and  $x_{1,n+1}\geq 0$.
}
\item{Transformation stage:
The second stage (T) of the DLM-based processor takes as input the
data stored in the six internal registers ${\bf Y}%
_{k,n+1}^{H}=( C_{k,n+1}^{H},S_{k,n+1}^{H}) $, ${\bf Y}%
_{k,n+1}^{V}=( C_{k,n+1}^{V},S_{k,n+1}^{V}) $, ${\bf Y}%
_{k,n+1}^{P}=( C_{k,n+1}^{P},S_{k,n+1}^{P}) $ and in the internal
vector ${\bf x}_{n+1}=( x_{0,n+1},x_{1,n+1}) $ and
combines the data into the eight-dimensional vector%
\begin{equation}
{\bf T}=
\left(
\begin{array}{c}
C_{0,n+1}^{H}C_{0,n+1}^{P}\sqrt{x_{0,n+1}} \\
S_{0,n+1}^{H}C_{0,n+1}^{P}\sqrt{x_{0,n+1}} \\
-S_{1,n+1}^{V}S_{1,n+1}^{P}\sqrt{x_{1,n+1}} \\
C_{1,n+1}^{V}S_{1,n+1}^{P}\sqrt{x_{1,n+1}} \\
C_{1,n+1}^{H}C_{1,n+1}^{P}\sqrt{x_{1,n+1}} \\
S_{1,n+1}^{H}C_{1,n+1}^{P}\sqrt{x_{1,n+1}} \\
-S_{0,n+1}^{V}S_{0,n+1}^{P}\sqrt{x_{0,n+1}} \\
C_{0,n+1}^{V}S_{0,n+1}^{P}\sqrt{x_{0,n+1}}
\end{array}%
\right) .
\end{equation}%
Rewriting the vector ${\bf T}$  as
\begin{eqnarray}
{\bf T}&=&
\left(
\begin{array}{c}
\left( C_{0,n+1}^{H}+iS_{0,n+1}^{H}\right) C_{0,n+1}^{P}\sqrt{x_{0,n+1}} \\
i\left( C_{1,n+1}^{V}+iS_{1,n+1}^{V}\right) S_{1,n+1}^{P}\sqrt{x_{1,n+1}} \\
\left( C_{1,n+1}^{H}+iS_{1,n+1}^{H}\right) C_{1,n+1}^{P}\sqrt{x_{1,n+1}} \\
i\left( C_{0,n+1}^{V}+S_{0,n+1}^{V}\right) S_{0,n+1}^{P}\sqrt{x_{0,n+1}}
\end{array}%
\right)
\nonumber \\
&\equiv&
\left(
\begin{array}{c}
a_{0}^{H} \\
ia_{1}^{V} \\
a_{1}^{H} \\
ia_{0}^{V}%
\end{array}%
\right) ,
\end{eqnarray}
shows that the operation performed by the transformation stage T corresponds
to the matrix-vector multiplication in the quantum theoretical description
of a PBS, namely%
\begin{equation}
\left(
\begin{array}{c}
b_{0}^{H} \\
b_{0}^{V} \\
b_{1}^{H} \\
b_{1}^{V}%
\end{array}%
\right) =
\left(
\begin{array}{cccc}
1 & 0 & 0 & 0 \\
0 & 0 & 0 & i \\
0 & 0 & 1 & 0 \\
0 & i & 0 & 0%
\end{array}%
\right) \left(
\begin{array}{c}
a_{0}^{H} \\
a_{0}^{V} \\
a_{1}^{H} \\
a_{1}^{V}%
\end{array}%
\right),
\end{equation}%
where $(a_{0}^{H},a_{0}^{V},a_{1}^{H},a_{1}^{V})$ and
$(b_{0}^{H},b_{0}^{V},b_{1}^{H},b_{1}^{V})$ denote the input and output
amplitudes of the photons with polarization $H$\ and $V$ in the 0 and 1
\revision{channels} of a PBS, respectively.
\revision{Note that in the quantum optical description of a (polarizing) beam splitter the vacuum
field must be included. In our simulation model, there is no need to introduce the concept of a vacuum field.}
}
\item{Output stage: The final stage (O) sends the message
\begin{equation}
{\bf w}=\left(
\begin{array}{c}
w_{0,n+1}/s_{0,n+1} \\
w_{1,n+1}/s_{0,n+1} \\
w_{2,n+1}/s_{1,n+1} \\
w_{3,n+1}/s_{1,n+1} \\
s_{0,n+1}/s_{2,n+1} \\
s_{1,n+1}/s_{2,n+1}%
\end{array}%
\right) ,
\end{equation}%
where
\begin{eqnarray}
w_{0,n+1} &=&C_{0,n+1}^{H}C_{0,n+1}^{P}\sqrt{x_{0,n+1}},\nonumber \\
w_{1,n+1} &=&S_{0,n+1}^{H}C_{0,n+1}^{P}\sqrt{x_{0,n+1}},\nonumber \\
w_{2,n+1} &=&-S_{1,n+1}^{V}S_{1,n+1}^{P}\sqrt{x_{1,n+1}},\nonumber  \\
w_{3,n+1} &=&C_{1,n+1}^{V}S_{1,n+1}^{P}\sqrt{x_{1,n+1}},\nonumber \\
s_{0,n+1} &=&\sqrt{w_{0,n+1}^{2}+w_{1,n+1}^{2}},\nonumber \\
s_{1,n+1} &=&\sqrt{w_{2,n+1}^{2}+w_{3,n+1}^{2}},\nonumber\\
s_{2,n+1} &=&\sqrt{w_{0,n+1}^{2}+w_{1,n+1}^{2}+w_{2,n+1}^{2}+w_{3,n+1}^{2}},\nonumber \\
&&
\end{eqnarray}
through output channel 0 if $w_{0,n+1}^{2}+w_{1,n+1}^{2}>r$ where
$0<r<1$ is a uniform pseudo-random number.
Otherwise, if $w_{0,n+1}^{2}+w_{1,n+1}^{2}\le r$, the output stage sends
through output channel 1 the message%
\begin{equation}
{\bf z}=\left(
\begin{array}{c}
z_{0,n+1}/t_{0,n+1} \\
z_{1,n+1}/t_{0,n+1} \\
z_{2,n+1}/t_{1,n+1} \\
z_{3,n+1}/t_{1,n+1} \\
t_{0,n+1}/t_{2,n+1} \\
t_{1,n+1}/t_{2,n+1}
\end{array}
\right),
\end{equation}
where
\begin{eqnarray}
z_{0,n+1} &=&C_{1,n+1}^{H}C_{1,n+1}^{P}\sqrt{x_{1,n+1}},\nonumber \\
z_{1,n+1} &=&S_{1,n+1}^{H}C_{1,n+1}^{P}\sqrt{x_{1,n+1}},\nonumber \\
z_{2,n+1} &=&-S_{0,n+1}^{V}S_{0,n+1}^{P}\sqrt{x_{0,n+1}},\nonumber \\
z_{3,n+1} &=&C_{0,n+1}^{V}S_{0,n+1}^{P}\sqrt{x_{0,n+1}},\nonumber \\
t_{0,n+1} &=&\sqrt{z_{0,n+1}^{2}+z_{1,n+1}^{2}},\nonumber \\
t_{1,n+1} &=&\sqrt{z_{2,n+1}^{2}+z_{3,n+1}^{2}},\nonumber \\
t_{2,n+1} &=&\sqrt{z_{0,n+1}^{2}+z_{1,n+1}^{2}+z_{2,n+1}^{2}+z_{3,n+1}^{2}}.\nonumber \\
&&
\end{eqnarray}
}
\end{itemize}
\revision{As mentioned earlier, the use of pseudo-random numbers to select the output channel
is not essential. We use pseudo-random numbers to mimic the apparent unpredictability
of the experimental data only. Instead of a uniform pseudo-random number generator, any algorithm
that selects the output channel in a systematic manner might be employed as well.
This will change the order in which messages are being processed but the content
of the messages will be left intact and the resulting averages do not change significantly.
}

\setlength{\unitlength}{1cm}
\begin{figure}[t]
\begin{center}
\includegraphics[width=8cm]{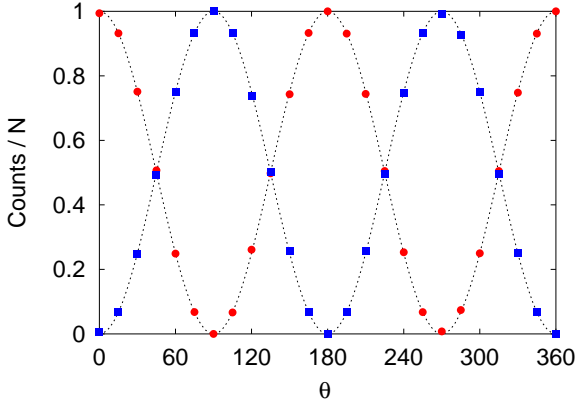}
\caption{Simulation results generated by the DLM network of the PBS shown in Fig.~\ref{PBS}.
Input channel 0 receives $(\cos\psi_0^H,\sin\psi_0^H,\cos\psi_0^V,\sin\psi_0^V,\cos\theta,\sin\theta)=(1,0,1,0,\cos\theta,\sin\theta)$.
Input channel 1 receives no events.
After each set of $N=10000$ events, $\theta$ is increased by $15^\circ$.
Squares and circles give the simulation results for the normalized intensities
$N_0/N$ and  $N_1/N$ as a function of $\theta$.
Dashed lines represent the results of quantum theory.
}
\label{malus}
\end{center}
\end{figure}

\setlength{\unitlength}{1cm}
\begin{figure}[t]
\begin{center}
\includegraphics[width=8cm]{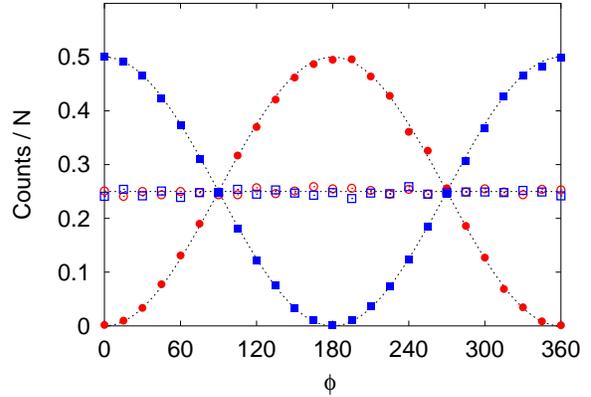}
\caption{Event-by-event simulation results of the experimental realization
of Wheeler's delayed-choice {\sl gedanken} experiment.
Open (closed) markers correspond to data for the open (closed) configuration of the interferometer.
Squares and circles give the simulation results for the normalized intensities $N_0/N$ and  $N_1/N$
as a function of the phase shift $\phi$ , $N_0$  ($N_1$)
denoting the number of events registered at detector D$_0$ (D$_1$). For each value of $\phi$,
the number of input events $N=10000$.
The number of detection events per data point is approximately the same as in experiment.
The simulation data is in qualitative agreement with experiment:
See Fig.~3 of Ref.~\cite{JACQ07}.
Dashed lines represent the results of quantum theory (Malus law).
}
\label{wave-particle}
\end{center}
\end{figure}

\subsection{Half wave plate (HWP)} This device performs a rotation of the
polarization of the photon~\cite{BORN64}. If the polarization of an incoming photon is at
an angle $\theta $ with respect to the optical axis of the HWP then, after
passing the HWP, the polarization of the photon has been rotated by an angle
$2\theta $. In order to change horizontal polarization into vertical
polarization, or vice versa, a HWP is used with its optical axis oriented at
$\pi /4$ . The HWP does not only change the polarization of the photon, but
also its phase as can be seen from the wave mechanical description%
\begin{equation}
\left(
\begin{array}{c}
b^{H} \\
b^{V}%
\end{array}%
\right) =\frac{-i}{\sqrt{2}}\left(
\begin{array}{cc}
\cos 2\theta & \sin 2\theta \\
\sin 2\theta & -\cos 2\theta%
\end{array}%
\right) \left(
\begin{array}{c}
a^{H} \\
a^{V}%
\end{array}%
\right) .
\end{equation}%
As a result, for the case $\theta =\pi /4$ , the polarization of the photon
is rotated by an angle $\pi /2$ and its phase is changed by $-\pi /2$.

\subsection{Electro-optic modulator (EOM)} This device rotates the polarization
of the photon by an angle depending on the voltage applied to the modulator.
In the laboratory experiment~\cite{JACQ07}, the EOM is operated such that
when a voltage is applied the EOM acts as a HWP that rotates the input
polarizations by $\pi /4$. In the simulation a pseudo-random number is used
to decide to apply a voltage to the EOM or not.
\revision{Also here we use a pseudo-random number to mimic the experimental procedure to control the EOM~\cite{JACQ07}.
Any other (systematic) sequence to control the EOM can be used as well.}

\subsection{Wollaston prism} This device is a PBS with one input channel and two output channels
and is simulated as the PBS described earlier. Messengers arrive at one and the same input channel only.

\subsection{Detection and data analysis procedure}Detector $D_0$ ($D_1$) registers
the output events at channel 0 (1).
During a run of $N$ events, the algorithm generates the data set
\begin{equation}
\Gamma =\left\{x_{n},A_{n}|n=1,...,N;\phi =\phi_{1}-\phi _{0}\right\} ,
\end{equation}
where $x_{n}=0,1$ indicates which detector fired ($D_{0}$ or $D_{1}$),
and $A_{n}=0,1$ is a pseudo-random number that is chosen
after the $n$th message (=photon) has passed the first PBS,
determining whether or not a voltage is applied to the
EOM (hence whether the configuration is open or closed).
The angle $\phi $ denotes the phase shift between the two interferometer arms.
This phase shift is varied by applying a plane rotation on the phase of the particles entering
channel 0 of the second PBS. This corresponds to tilting the second PBS in
the laboratory experiment~\cite{JACQ07}. For each phase shift $\phi $ and
interferometer configuration (open or closed) the number of 0 (1) output
events $N_{0}$ ($N_{1}$) is calculated.

\section{Simulation results}
The algorithm described above directly translates
into a simple computer program.
For a fixed set of input parameters, each simulation takes a few seconds on a
present-day PC. All \revision{simulations} are carried out with $\alpha=0.99$~\cite{RAED05d}.

We first demonstrate that our model for the PBS reproduces Malus' law.
In this simulation, we send messengers to one input channel, say channel 0, only.
This implies that the registers that are connected to channel 1 will not change
during the simulation.
Figure~\ref{malus} shows a representative set of event-based
simulation results for the PBS modeled by the processor depicted in Fig.~\ref{PBS}. The data set
is produced with input messages $\left( \cos \psi _{0}^{H},\sin \psi
_{0}^{H},\cos \psi _{0}^{V},\sin \psi _{0}^{V},\cos \xi _{0},\sin \xi
_{0}\right) $. The values of $\psi _{0}^{H}$ and $\psi _{0}^{V}$\ are fixed
but irrelevant otherwise. Also the value of $\xi _{0}$ is
irrelevant. It is clear that the intensities in both output channels obey
Malus' law.

Next, we build a network that contains all the optical
components of the laboratory experiment~\cite{JACQ07} (see Fig.~\ref{wheeler}).
Before the simulation starts we set ${\bf x}%
_{0}=( x_{0,0},x_{1,0}) =( r,1-r) $, where $r$ is a
uniform pseudo-random number. In a similar way we use pseudo-random numbers
to initialize ${\bf Y}_{0,0}^{H}$, ${\bf Y}_{0,0}^{V}$, $%
{\bf Y}_{0,0}^{P}$, ${\bf Y}_{1,0}^{H}$, $%
{\bf Y}_{1,0}^{V}$ and ${\bf Y}_{1,0}^{P}$.
In this simulation, we send messengers to one input channel
(see Fig.~\ref{wheeler}).
The HWP in BS$_{output}$ changes the phases and also interchanges the roles of channels 0 and 1.
Disregarding a few exceptional events, the PBS in BS$_{output}$ generates messages in one of the channels only.

Representative results of an event-by-event simulation of this network, a one-to-one image of
Wheeler's delayed choice experiment~\cite{JACQ07}, are shown in Fig.~\ref{wave-particle}.
The simulation data are in quantitative agreement with the averages calculated from quantum theory
and in qualitative agreement with experiment~\cite{JACQ07}.

\section{Conclusion}

In this Letter, we have proven that it is possible to give a particle-only description
for both the open and closed interferometer configuration in Wheeler's delayed choice
experiment that (1) reproduces the averages calculated from quantum theory,
(2) satisfies Einstein's criteria of realism and local causality,
(3) does not rely on any concept of quantum theory \revision{or of probability theory},
and (4) is not in conflict with common sense.

\bibliographystyle{eplbib}
\bibliography{../../epr}

\end{document}